\documentclass[aps,pra,twocolumn,amsmath,amssymb,nofootinbib,superscriptaddress]{revtex4}
\usepackage[english]{babel}
\usepackage{latexsym}
\usepackage{graphics}
\usepackage{epsfig}
\usepackage{color}
\usepackage{bm}
\usepackage{nicefrac}
\usepackage{amsmath}
\usepackage{amssymb}
\usepackage{array,multirow}
\usepackage[normalem]{ulem}
\usepackage{color}
\usepackage{siunitx}

\newcommand{\be}{\begin{equation}}
\newcommand{\ee}{\end{equation}}
\newcommand{\bra}{\langle}
\newcommand{\ket}{\rangle}
\newcommand{\Rvec}{\ensuremath{\boldsymbol{R}}}
\newcommand{\rvec}{\ensuremath{\boldsymbol{r}}}
\newcommand{\kvec}{\ensuremath{\boldsymbol{k}}}
\newcommand{\xvec}{\ensuremath{\boldsymbol{x}}}
\newcommand{\yvec}{\ensuremath{\boldsymbol{y}}}

\begin{document}

\title{Universal Short Range Correlations in Bosonic Helium Clusters}
\author{B. Bazak}
\affiliation{The Racah Institute of Physics, The Hebrew University, 9190401,
  Jerusalem, Israel}
\author{M. Valiente}
\affiliation{Institute for Advanced Study, Tsinghua University, Beijing 100084, China}
\author{N. Barnea}
\affiliation{The Racah Institute of Physics, The Hebrew University, 9190401,
  Jerusalem, Israel}

\date{\today}

\begin{abstract}
  Short-range correlations in bosonic Helium clusters, composed of $^4$He atoms,
  are studied utilizing the generalized contact formalism. The emergence of
  universal $n$-body short range correlations is formulated
  and demonstrated numerically via Monte Carlo simulations. The values of
  the $n$-particle contacts are evaluated for $n\le5$. In the thermodynamic
  limit, the two-body contact is extracted from available experimental
  measurements of the static structure factor of liquid $^4$He at high momenta,
  and found in a good agreement with the value extracted from our calculations. 
\end{abstract}

\maketitle

Interacting multiparticle systems where the interaction range is much smaller
than any other characteristic length scale, such this associated with the
density or the average momentum, can be studied using the zero range
approximation. In this limit, the interaction details are neglected
and the system acquires universal features depending only on its density $\rho$
and the scattering length $a_s$.
When $a_s$ is small, the particles interact weakly and 
the system is amenable to perturbative treatment. 
When it is large, the particles are strongly correlated and one
needs to resort to numerical methods to study the system properties.

About a decade ago, while studying two-component Fermi systems with
large $a_s$, Tan has succeeded to show that many of its properties 
are governed by a single parameter, the so-called \emph{contact} $C$, which
measures the probability of two particles being in close proximity \cite{Tan08}.
Following Tan's work, different relations between various properties of such
system and the contact, known as the \emph{Tan relations}, were derived and
verified experimentally with ultracold gases
\cite{SteGaeDra10,SagDraPau12,WerTarCas09,KuhHuLiu10}.
One example is the one-body momentum distribution $n(k)$ tail, determined to be
\be \label{tail}
\lim_{k \to \infty} n(k) = C/k^4.
\ee

The Pauli principle prevents identical fermions from approaching
each other in a relative $s$-wave state. Consequently, three-body correlations
are typically negligible in an ultracold two-component atomic Fermi gas.
In contrast, such three-body coalescence is expected to play a
decisive role in bosonic gases or for
nucleons, where the spin-$\frac{1}{2}$ neutrons and protons
form a four-component Fermi system.
Indeed for bosonic systems, the tail of the momentum distribution is predicted
to include a subleading $k^{-5}$ term, emerging from such three-body
correlations \cite{BraKanPla11}.
We note that other singular interactions, like Coulomb, also exhibit universal
short-distance correlations \cite{Kim73,HofBarZwe13}.

To derive the {\it Tan relations} one may start with the observation
that when two particles approach each other, 
the $N$-body wavefunction is factorized into a product of a 
universal 2-body function $\phi_2$ and 
a state-dependent function $A_N^{(2)}$ describing the residual system,
\begin{equation}\label{factorization2}
  \Psi(\rvec_1,\ldots,\rvec_N)\xrightarrow[r_{ij} \to 0]{} \phi_2(\rvec_{ij})
  A_2^{(N)}(\Rvec_{ij},\{\rvec_k\}_{k\ne i,j}) {.}
\end{equation}
Here $\rvec_{ij}=\rvec_i-\rvec_j$ is the interparticle distance and
$\Rvec_{ij}=(\rvec_i+\rvec_j)/2$ is the pair's center of mass coordinate.
In the zero-range approximation the universal pair wave-function is given by
$\phi_2(\rvec_{ij})=1/r_{ij}-1/a_s + O(r_{ij})$.

Recently, the contact formalism was generalized to systems where the zero-range
approximation is not justified
\cite{WeiBazBar15a,WeiBazBar15b,WeiCruBar18,Mil18,WerCas12ref}.
This is the situation, for example, in the atomic nucleus, where
the interparticle distance is about $2.4\,\mathrm{fm}$,
while the nuclear interaction range, estimated from the
pion mass, is about $\hbar/m_{\pi}c\approx 1.4\,\mathrm{fm}$.
This is also the situation in $^4$He atomic clusters, where the average
interparticle distance within clusters with more than three atoms is about 
$5\,\mathrm{\AA}$, while the van der Waals length,
characterizing the potential's range, is about $5.4\,\mathrm{\AA}$.

In such cases, one would not expect to see a \emph{strong} universality,
i.e. relations which do not depend on the interaction details
and are determined only by scattering parameters such as $a_s$. 
Still, given an interaction model strong at small distances, 
the wavefunction factorization
(Eq.~\ref{factorization2}) 
remains valid since at close distance a correlated particle pair is
barely influenced by the surrounding particles and therefore its wavefunction
$\phi_2(\rvec)$ should be the same regardless of the system size or state.
We will call this situation \emph{weak} universality.

It is instructive therefore to study the adaptation of Tan's relations to
weak universality. For instance, relations between the one and two-body momentum
distributions as well as the two-body density were studied in nuclei
\cite{WeiBazBar15b, WeiCruBar18}. 
In the following we will investigate such relations for bosonic $^4$He clusters.

$^4$He clusters have attracted a lot of attention.
For a long time, the $^4$He trimer seemed to be the most promising candidate
for experimental validation of the Efimov effect \cite{Efimov70}, as liquid
Helium was for Bose-Einstein condensation. Recently $^4$He dimer and trimer
densities were measured experimentally \cite{Zeller16,Voi14}.
The results compare very well with theoretical calculations using
$^4$He pair potential models.
The dimer and trimer densities at short range play a crucial role in the
contact formalism we study here. 
The atomic clusters exhibit a universal short-range $2,3$-body behavior stemming
from the dimer and trimer wavefunctions, respectively. Moreover, this phenomenon
also continues with the coalescence of more atoms inside these clusters,
showing the emergence of $4,5,\ldots$-body universality.

In order to study the properties of $^4$He clusters we solve the $N$-body
Schr\"odinger equation with the LM2M2 pair potential \cite{AziSla91}.

As we argued above, in the limit of vanishing interparticle distance $r\to 0$
we expect the wavefunction $\Psi$ to factorize as in
Eq. (\ref{factorization2})
into a universal $2$-body function and a residual state dependent function. 
If true, this factorization holds also for $N=2$. Consequently, we can
identify $\phi_2$ with the dimer wavefunction.

The resulting two-body contact is defined as the
norm of the residual non-universal part of the wavefunction
multiplied by the number of pairs,
\be\label{def_c2}
   C_2^{(N)}=\frac{N(N-1)}{2}\bra A_2^{(N)} | A_2^{(N)} \ket
           =\binom{N}{2}\bra A_2^{(N)} | A_2^{(N)} \ket.
\ee
Using this definition, the pair density function at short distances
attains an extremely simple form,
\be \label{rho_2_N}
  \rho_2^{(N)}(r) = \bra \Psi | \hat \rho^{(N)}_2(r) | \Psi \ket
      \xrightarrow[r \to 0]{} 
    C_2^{(N)} \rho_2(r)
\ee 
where
$\hat \rho^{(N)}_2(r) = \frac{1}{r^2} \sum_{i<j} \delta(r_{ij}-r)$,
$\rho_2(r) \equiv \rho_2^{(2)}(r) = \int d\Omega_2 |\phi_2(\rvec)|^2$,
and $\Omega_2$ is the solid angle.

In a bosonic system, coalescence of more particles should 
provide further factorizations of the wavefunction \cite{WerCas12}.
When particles $i,j$ and $k$ come close together, the wavefunction is
factorized as
\be\label{factorization3a}
  \Psi \xrightarrow[r_{ijk} \to 0]{}
            \phi_3(\xvec_{ijk},\yvec_{ijk})
            A_3^{(N)}(\Rvec_{ijk},\{\rvec_l\}_{l\ne i,j,k})
\ee
where the triplet wavefunction depends on the Jacobi coordinates
$\xvec_{ijk}=\sqrt{1/2}(\rvec_i-\rvec_j)$ and
$\yvec_{ijk}=\sqrt{2/3}(\rvec_k-(\rvec_i+\rvec_j)/2)$,
and the factorization holds for small hyperradius
$r_{ijk}^2=x_{ijk}^2+y_{ijk}^2$.
Here $\Rvec_{ijk}$ is the three body center of mass coordinate.
In analogy with Eq. \eqref{def_c2},
the three-body contact in the $N$-body system is defined to be the
number of triplets times the norm of the particular part of the wavefunction
in three-body coalescence, 
\be\label{def_c3}
    C_3^{(N)}=\binom{N}{3}\bra A_3^{(N)} | A_3^{(N)} \ket.
\ee
The triplet density operator is defined as,
\be 
  \hat \rho^{(N)}_3(r) = \frac{1}{r^5} \sum_{i<j<k} \delta(r_{ijk}-r)
\ee
and its expectation value in the $N$-body system is
\be \label{rho_3_N}
   \rho^{(N)}_3(r) = \bra \psi | \hat \rho^{(N)}_3(r) | \psi \ket
               \xrightarrow[r \to 0]{} 
               C_3^{(N)} \rho_3(r)
\ee
where
$\rho_3(r) \equiv \rho_3^{(3)}(r) = \int d\Omega_3 \; |\phi_3(\xvec,\yvec)|^2$,
and $\Omega_{3}$ denotes the hyperangles associated with $\xvec$, and $\yvec$.

Similar factorization is assumed in the $n$-body coalescence, leading to the
definition of the $n$-body contact, and to the $n$-body density function,
\be\label{rho_n_N}
   \rho^{(N)}_n(r) \xrightarrow[r \to 0]{} C_n^{(N)} \rho_n(r), 
\ee
where here $r=\sqrt{\sum_{i < j}^n(\rvec_i-\rvec_j)^2/n}$ is the $n$-body
hyperradius. 
This is one of the main results of this paper and in the following we shall
show that this is indeed the case for $n \leq 5$ in atomic $^4$He droplets
with $N$ atoms. 
In the mean time we note that with the above definition the contact for $n=N$
equals unity since $\rho_n(r)\equiv \rho^{(n)}_n(r)$.

Using this factorization, the zero-range result for the high momentum limit
of the 1-body momentum distribution (Eq.~\ref{tail}), is now modified to get
\cite{WerCas12a}
\be\label{momentum_distrib_1b}
   n^{(N)}(\kvec) \xrightarrow[k\to\infty]{} 2 C^{(N)}_2 |\tilde\phi_2(\kvec)|^2
\ee 
where $\tilde\phi_2(\kvec)$ is the Fourier transform (FT) of 
$\phi_2(\rvec)$. 
The high momentum limit of the static structure factor, which is proportional 
to the contact in the zero-range limit \cite{KuhHuLiu10}, gets now the form
\be\label{structure_factor}
  S(Q) \xrightarrow[Q\to\infty]{} 1+\frac{2C_2^{(N)}}{N}\frac{4\pi}{Q}\int dr r
  \sin(Q r) \rho_2(r)\;,
\ee
where $Q$ is the momentum transfer. 
It is also possible to relate the contact to the potential energy which,
for a cluster of bosons interacting via 2-body forces
can be written using the 2-body density
$\bra V_2^{(N)} \ket = \int d\rvec \rho_2^{(N)}(r)v(r)$.
For a short range interaction we can replace $\rho_2^{(N)}$ by its asymptotic
form, Eq. \eqref{rho_2_N}, relating the $N$-body
potential energy to the 2-body contact
and potential energy \cite{WerCas12a},
\be \label{potential}
   \bra V_2^{(N)} \ket = C_2^{(N)} \bra V_2^{(2)} \ket
\;,
\ee
which generalizes the zero-range result of Ref. \cite{ZhaLeg09}.

\emph{The $N$ dependence -}
To understand the dependence of the $n$-body contact
on the total particle number $N$ in the cluster, it is useful to 
relate the pair density $\rho_2^{(N)}$ 
to the $2$-body density 
   $\chi(\rvec,\rvec') = \sum_{i\neq j}\bra \Psi|\delta(\rvec-\rvec_i)\delta(\rvec'-\rvec_j)|\Psi\ket$,
namely
\be
    \rho_2^{(N)}(\rvec_{12}) = \frac{1}{2}\int d\Rvec_{12}\chi(\rvec_1,\rvec_2) \;.
\ee
In the limit $N\to\infty$ the system becomes homogeneous,
$\chi(\rvec_1,\rvec_2) \to \chi(\rvec_{12}) $ and therefore
$\rho_2^{(N)}(\rvec_{12}) = V\chi(\rvec_{12})/2
                        = N\chi(\rvec_{12})/2\rho$
where $V$ is the volume of the system and $\rho=N/V$ is the density. Taking now
the limit $r_{12}\to 0$ one can get the relation \cite{WerCas12a}
\be\label{chi_r}
    \chi(r)  \xrightarrow[r \to 0]{} 2\rho\frac{C_2^{(N)}}{N}\rho_2(r) \;.
\ee
We know that in the thermodynamic limit  $\chi$ and $\rho$ are finite.
It follows that $C_2^{(N)}\propto N$ as $N\to\infty$. The same argument can be
repeated for $n=3,4,5,\ldots$ leading to the general conclusion that for any
$n$-body coalescence $C_n^{(N)}\propto N$ as $N\to\infty$.
Equipped with this observation it seems natural to define a reduced contact
$\tilde{C}_n^{(N)} \equiv C_n^{(N)}/N$. As the atomic He clusters behave very
much like a cluster of rigid balls, 
we expect that the leading corrections to the above argument will depend on the
ratio between surface particles $\propto N^{2/3}$ and volume particles
$\propto N$.
Consequently in the limit $N\to\infty$ the contacts are expected to have the
following $N$ dependence
\be\label{c_asymptot}
   \tilde{C}_n^{(N)}=\tilde{C}_n^{\infty}+\alpha_nN^{-1/3}+\beta_nN^{-2/3}+\ldots
\ee 
\emph{The computational method -}
Throughout the years, a variety of numerical methods have been developed to
solve the few-body Schr\"odinger equation.
However, the increasing dimensionality and the
hard-core nature of the $^4$He-$^4$He pair potential make this problem 
hard to handle for most numerical methods.
Here we use the Variational Monte Carlo (VMC) and Diffusion Monte Carlo (DMC)
methods.
Since these methods are well-known we will only describe them very briefly, 
for a comprehensive review see e.g. \cite{KalWhi08}.

Given a trial wave-function $\Psi_T$, the variational energy 
\be \label{var}
   E_{var}=\frac{\bra \Psi_T | H | \Psi_T \ket}{\bra \Psi_T | \Psi_T \ket}\ge E_0
\ee
is an upper bound to the true ground-state energy $E_0$. In the VMC method
the integrals in
Eq.~\eqref{var} are evaluated using the Monte Carlo numerical integration
technique, typically the Metropolis algorithm \cite{MetRosRos53}.
Using the variational principle \eqref{var}, parameters characterizing
$\Psi_T$ can be optimized, minimizing the trial energy or its variance.

DMC is an alternative approach to solve the
Schr\"odinger equation through propagation of the solution
in imaginary time $\tau = -it$,
\be \label{DMC}
   \frac{\partial \Psi(\rvec_1\ldots \rvec_N,\tau)}{\partial \tau} =
   \left(T + V - E_R \right) \Psi(\rvec_1\ldots \rvec_N,\tau).
\ee
where $E_R$ is a reference energy.
Eq. (\ref{DMC}) is treated as a diffusion-reaction process for so-called
walkers, distributed according to $\Psi$.
As time propagates, $\Psi$ will be dominated by the eigenstate with the
lowest energy which has a non-zero overlap with the initial state.
All other eigenstates will decay exponentially faster.
The ground state energy is the reference energy which conserves the
walkers number.

Improved results are obtained by introducing a trial wavefunction to guide
the diffusion process, therefore a typical DMC 
calculation starts with an optimized VMC wave-function.
We adopt the trial wavefunction form of Ref. \cite{RicLynDol91}, 
$\Psi_T=\prod_{i<j}f(r_{ij})$
where
\be
  f(r)=\exp\left[-(p_{5}/r)^5-(p_{2}/r)^2- p_1 r \right]/r^{p_0}.
\ee
Here $p_5,p_2,p_1,$ and $p_0$ are variational parameters, which can be found
in Ref. \cite{SM}. 

\emph{Ground state energies - }
To benchmark our Monte Carlo code we have calculated the ground-state energies
of small $^4$He clusters with the LM2M2 pair-potential.
Calculations were done with $4000$ walkers, using $10000$ blocks of $500$
iterations each. The first $100$ blocks were used for equilibration.

The $^4$He trimer ground state energy using this potential has been
calculated using several few-body techniques. 
Most results agree with $B_3 = 126.0(5)$ mK
\cite{NieFedJen98,BluGre00,RouYak00,MotSanSof01,LazCar06,GuaKorNav06,SalYarLev06,HiyKam12},
while different values also exist \cite{EsrLinGre99,GonRubMir99}.

Few calculations have been done for larger clusters. The tetramer energy was
calculated in Refs. \cite{BluGre00,LazCar06,GuaKorNav06,HiyKam12}
using the LM2M2 potential.
In Ref.~\cite{GatKieViv11} a soft-core potential was used
while in Refs.~\cite{PlaHamMei04,BazEliKol16} an effective
field theory approach was followed. In both cases the interaction parameters
were fitted to the LM2M2 potential.
Larger clusters were investigated using the DMC method
~\cite{BluGre00,GuaKorNav06}.
In Table \ref{tbl:Energies} we compare these calculations with our results,
showing good agreement with the published binding energies.

\begin{table}
\begin{center}
  \caption{The ground-state energies (in mK) of small $^4$He clusters, with the
    LM2M2 pair-potential. The dimer energy is $1.30348$ mK \cite{HiyKam12}.
    \label{tbl:Energies}}
\vspace{0.3cm}
       {\renewcommand{\arraystretch}{1.25}%
\begin{tabular}
{c@{\hspace{3mm}} c@{\hspace{4mm}}  c@{\hspace{4mm}} c@{\hspace{5mm}} c@{\hspace{5mm}} c}
\hline\hline 
$N$ & Ref \cite{LazCar06} & Ref \cite{HiyKam12} & Ref \cite{BluGre00} & Ref \cite{GuaKorNav06} & This work \\
\hline
3  & 126.39 & 126.40 & 125.5(6)& 124(2)  & 125.9(2) \\
4  & 557.7  & 558.98 & 557(1)  & 558(3)  & 557.4(4) \\
5  &        &        & 1296(1) & 1310(5) & 1300(2)  \\
6  &        &        & 2309(3) & 2308(5) & 2315(2)  \\
7  &        &        & 3565(4) & 3552(6) & 3571(2)  \\
8  &        &        & 5020(4) & 5030(8) & 5041(2)  \\
9  &        &        & 6677(6) & 6679(9) & 6697(2)  \\
10 &        &        & 8495(7) & 8532(10)& 8519(3)  \\
\hline\hline
\end{tabular}}
\end{center}
\end{table}

\emph{The $n$-body density function --}
To calculate the $n$-body densities we have
used a combination of VMC and DMC estimates,
\be
 \bra \hat O \ket = 2 \bra \hat O \ket_\textrm{DMC} 
                    - \bra \hat O \ket_\textrm{VMC}
\ee
where 
$\bra \hat O \ket_\textrm{DMC} =
\bra \Psi_T | \hat O | \Psi \ket/\bra \Psi_T | \Psi \ket$ 
is the mixed DMC estimate, and
$\bra\hat O \ket_\textrm{VMC}=\bra\Psi_T|\hat O|\Psi_T\ket/\bra\Psi_T|\Psi_T\ket$
is the VMC estimate. This result is accurate to second order 
in the wavefunction $O(\delta \Psi^2)$, $\delta \Psi = \Psi_T - \Psi$ 
\cite{CeperlyKalos1979}.
Moreover, we checked our results with pure-estimator based on the descendant weighting
method \cite{Kal70}, and found no significant change.

For the smaller clusters, the resulting $n$-body densities exhibit a typical
bell shape, starting from zero at $r=0$, reaching a maximum value,
and finally falling exponentially at large $r$.
According to Eq. (\ref{rho_2_N}), we expect that at short distances 
the pair density function $\rho_2^{(N)}$ will coincide with the dimer density
$\rho_2$ up to a scaling factor, the 2-body contact $C_2^{(N)}$,
which we can extract fitting these two functions \cite{extract_contacts}.
This situation is expected to repeat itself for the 3-body density function, 
Eq. \eqref{rho_3_N},
and in general for any $n$-body density, Eq. \eqref{rho_n_N}.

\begin{figure}
\begin{center}
\includegraphics[width=0.5\textwidth]{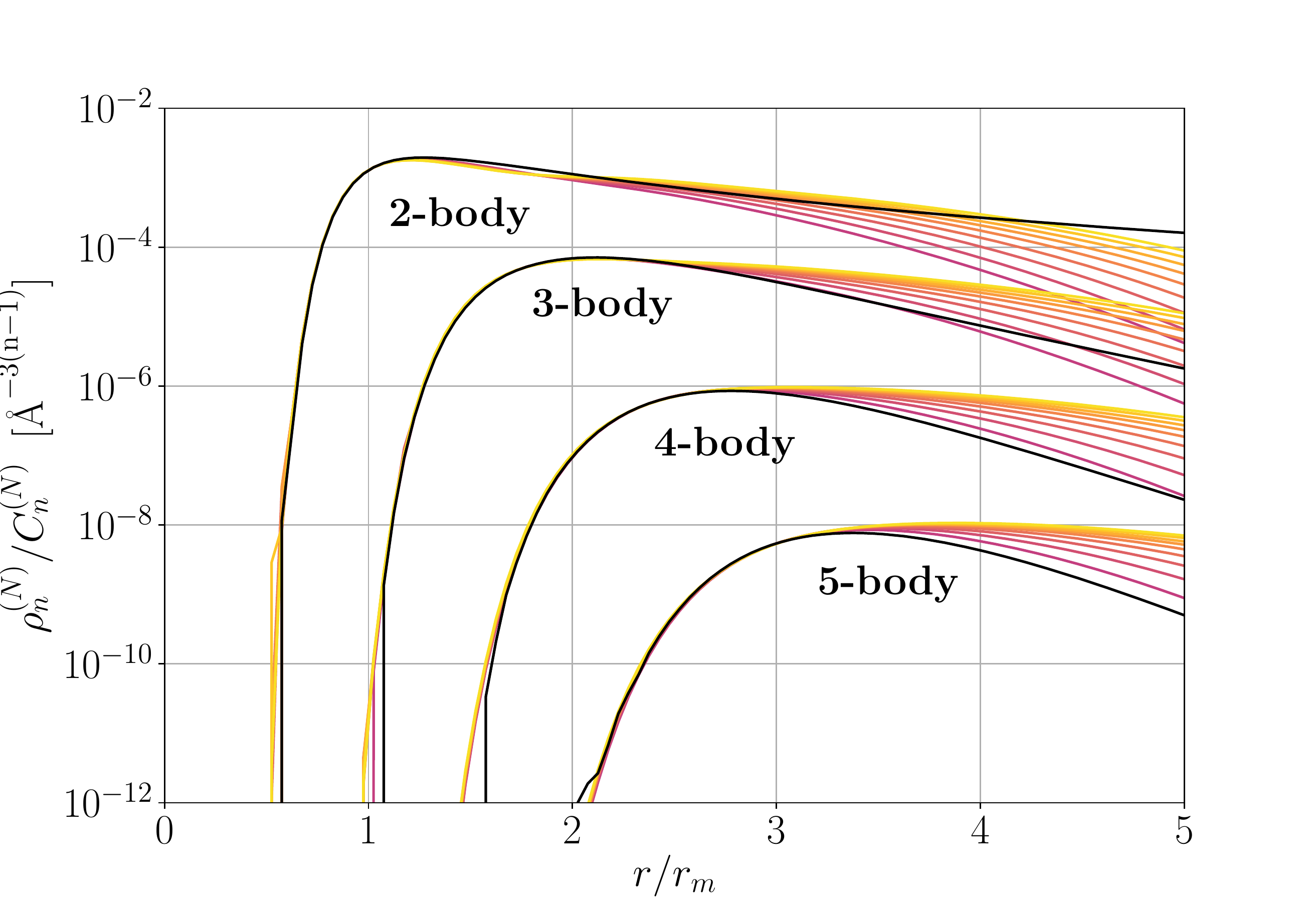}
\caption{\label{fig:n_body_density}
  The $n$-body density function normalized with the 
  appropriate contact $\rho_n^{(N)}/C_n^{(N)}$ is
  presented as function of the $n$-body radius for $n=2,3,4,5 $. 
  For each $n$ the reference density $\rho_n$ is drawn with a black line.
  The densities for $N=10, 15, 20 \ldots 50$, are given by the colored
  lines (from dark to light).}
\end{center}
\end{figure}

Having extracted the contacts \cite{extract_contacts} we are in position
to demonstrate the validity of Eq. \eqref{rho_n_N}. To this end, 
we plot in Fig. \ref{fig:n_body_density}
the normalized $n$-body densities $\rho_n^{(N)}/C_n^{(N)}$ as a function of the
$n$-body radius $r/r_m$. $r_m = 2.6965 \rm{\AA}$ being the minimum 2-body
potential locus. The plot contains results for $^4$He 
clusters with $N=n$ and $N=10,15,20,\ldots 50$ particles.
Inspecting the plot we see that, indeed, for each $n$ there is a range $r_n$  
such that for $r \leq r_n$ all the normalized densities collapse into a single
curve. For the pair density this range is approximately $ 1.3 r_m$ and it grows
linearly with $n$, i.e. $r_n \approx n \; 0.65 r_m$.

\begin{figure}
\begin{center}
\includegraphics[width=0.5\textwidth]{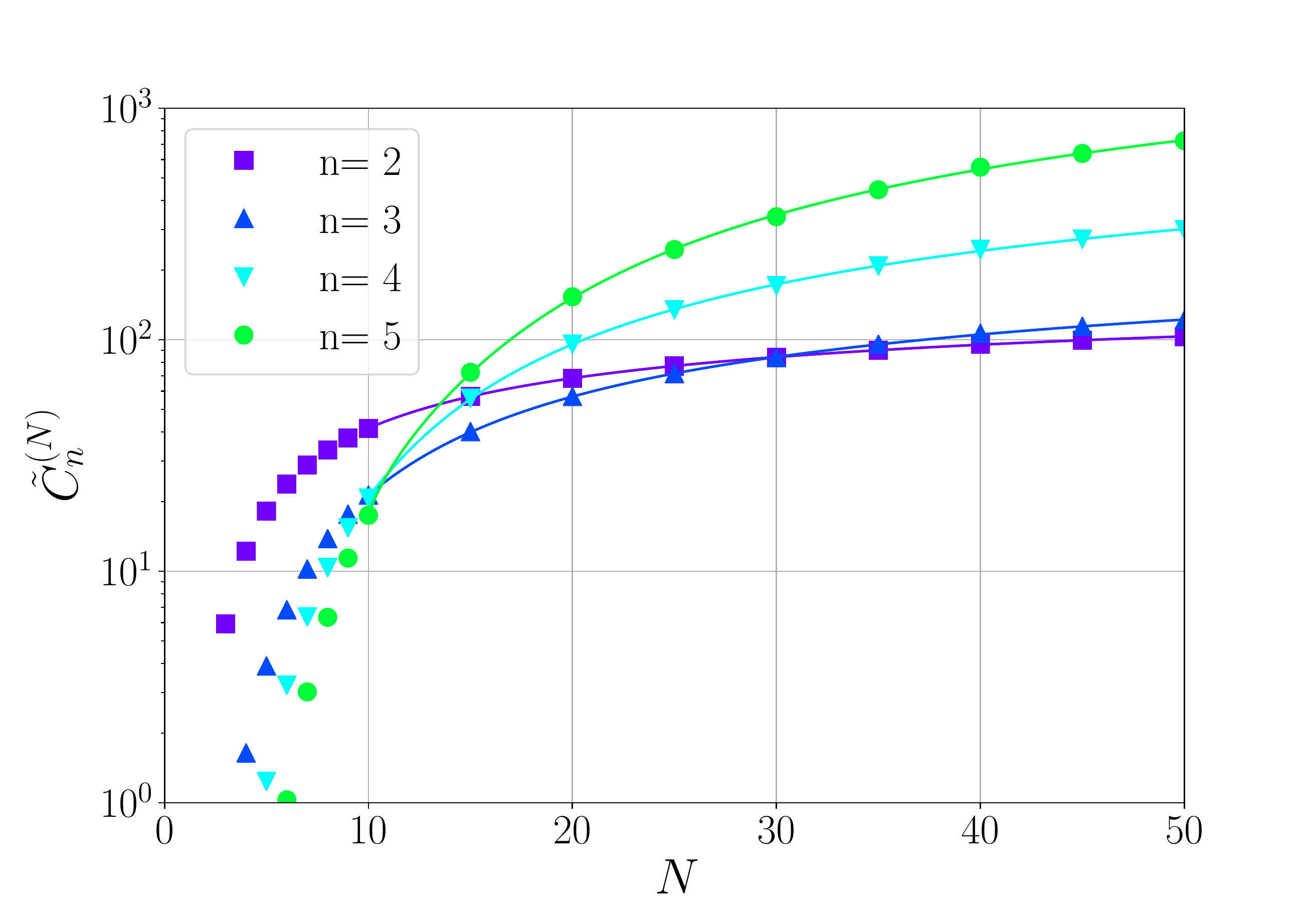}
\caption{\label{fig:Contacts_nn}
  The evolution of the reduced $n$-body contacts $\tilde{C}_n^{(N)}={C}_n^{(N)}/N$
  with the system size $N$. Symbols - calculated values, curves - the asymptotic
  expansion given in Eq. \eqref{c_asymptot}.}
\end{center}
\end{figure}

The numerical values of the extracted contacts are
presented in the supplementary material \cite{SM}. Here we analyze
the $N$ dependence of the $n$-body contacts. From Eq.~\eqref{c_asymptot}
we expect $\tilde C_n^{(N)}=C_n^{(N)}/N$ to be finite in the thermodynamic limit.
Our MC code was designed to study small He clusters with $N\leq 50$ particles, 
and is therefore ill-equipped to study this $N\to\infty$ limit.
Instead, to estimate $\tilde C_n^{\infty}$ we fit our calculated contacts
to Eq. \eqref{c_asymptot}. Doing so, we have found
that, for $N \geq 10$, $3$ terms are enough to describe $C_2^{(N)}, C_3^{(N)}$, 
 and $4$ terms for $C_4^{(N)}, C_5^{(N)}$.
The asymptotic values of the reduced contacts are given in Table
\ref{tbl:Contacts}.
The calculated contacts are plotted together with the
asymptotic expansion in Fig. \ref{fig:Contacts_nn}, where we observe that the
calculated values are well reproduced by the asymptotic expansion. 

\begin{table}
  \caption{The asymptotic values of the reduced $n$-body contacts 
           $\tilde{C}_n^{(N)}={C}_n^{(N)}/N$ of $^4$He droplets. 
    \label{tbl:Contacts}}
\begin{center}
\begin{tabular}
{c@{\hspace{6mm}} c@{\hspace{4mm}}  c@{\hspace{4mm}} c@{\hspace{4mm}}  c@{\hspace{4mm}} }
\hline\hline 
 $n$   & 2 & 3 & 4 & 5 \\
\hline
$\tilde{C}_n^{\infty}$ &
    $230 \pm 25$   &  
    $500 \pm 60$   &  
    $1800 \pm 300$ &  
    $5900 \pm 1000$ \\       
\hline\hline
\end{tabular}
\end{center}
\end{table}

Having calculated the 2-body contacts, the $Q\to\infty$ limit of the
structure factor can be evaluated for any helium droplet and compared with
experiment. 

\begin{figure}
\begin{center}
\includegraphics[width=0.5\textwidth]{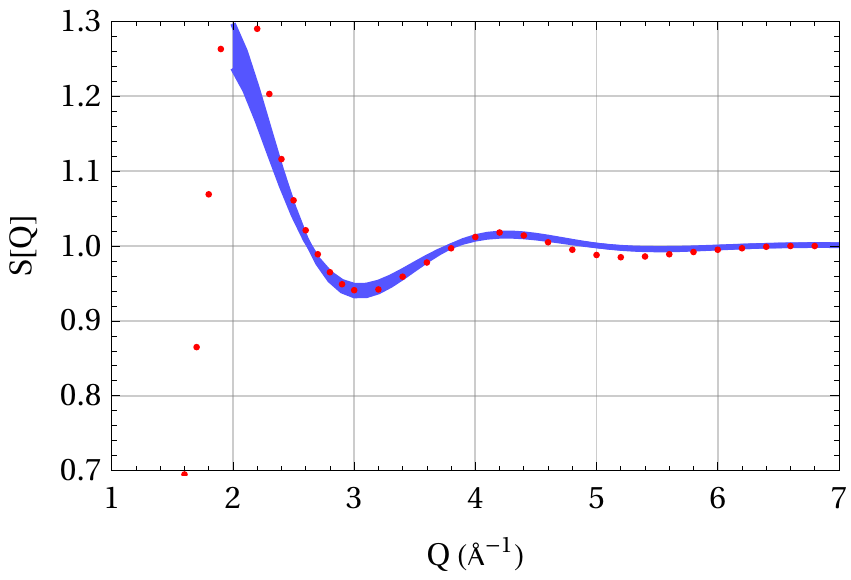}
\caption{\label{fig:SF}
  The structure factor of liquid $^4$He as a function of the momentum
  transfer $Q$, a comparison between the experimental data
  of Svensson {\it et al.} \cite{SveSeaWoo80} and the contact theory,
  Eq. \eqref{structure_factor}. The experimental data are presented by dots.
  The band corresponds to calculated contact values in the range
  $\tilde{C}_2^\infty\in (200,250)$.
}
\end{center}
\end{figure}

For liquid helium, the structure factor was measured using x-ray scattering
\cite{RobHal82,WirHal87}, and neutron scattering techniques \cite{SveSeaWoo80}.
Following the analysis of Donnelly and Barenghi \cite{DonBar98} we adopt the 
latter data set and compare it  with the contact theory, Fig. \ref{fig:SF}.
In the range $Q\geq 2 \rm{\AA}^{-1}$, 
dominated by the short-range pair function $\phi_2$, 
we see a nice agreement between the two.
The data fits contact values in the range
$\tilde C_2^\infty\in (200,250)$ as predicted by our calculations, 
Tab. \ref{tbl:Contacts}.

The dynamic structure factor $S(Q,E)$ of liquid $^4$He was recently measured
by Prisk {\it et al.} \cite{Prisk2017}, using the neutron Compton scattering
technique. In the impulse approximation, $S(Q,E)$ and consequently the neutron
Compton profile can be calculated from the 1-body momentum distribution
$n(\kvec)$. Utilizing the contact relation \eqref{momentum_distrib_1b},
we analyzed these results. Overall we got reasonable agreement between the
data and the theory for contact values $\tilde{C}_2^\infty= 180 \pm 40$, a
value consistent with both the MC calculation and the static structure
factor data. 

\emph{Conclusion.}
  Summing up, utilizing the generalized contact formalism, we have studied
  short-range correlations in bosonic Helium clusters composed of $^4$He
  atoms. Specifically, we have studied $n$-body coalescences, and
  the emergence of universal $n$-body short-range correlations.
  Employing the LM2M2 pair potential, VMC and DMC calculations were used to
  demonstrate and verify the universal nature of these correlations.
  For systems with up to $N=50$ particles, the values of the $n$-body contacts
  were evaluated numerically for $n\le 5$. 
  The thermodynamic limit was studied, extrapolating our numerical results. 
  Comparing our prediction with the experimental two-body
  contact, extracted from available measurements of the structure factor of
  liquid $^4$He at high momenta, we have found a good agreement.
  It would be interesting to compare our predictions with detailed 
  Monte Carlo simulations of Helium liquid.

  The implications of the current formalism on the momentum distribution and
  the dynamic structure factors call for further experimental studies in the
  high momentum sector.

\begin{acknowledgments}
  We would like to thank Reinhard D\"orner, Gregory Astrakharchik, 
  Dmitry Petrov, Lorenzo Contessi and Ronen Weiss
  for useful discussions and communications.
  We thank Timothy R. Prisk for sharing the experimental data of
  Ref. \cite{Prisk2017} with us.
  The work of N.B was supported by the Pazy foundation.
\end{acknowledgments}

\onecolumngrid

\setcounter{table}{0}
\renewcommand{\thetable}{S\arabic{table}}

\vspace{\columnsep}

\newpage
\begin{center}
\textbf{\large Supplemental Material: Universal short range correlations in bosonic Helium clusters}
\end{center}

\setcounter{equation}{0}
\setcounter{figure}{0}
\setcounter{page}{1}
\makeatletter
\renewcommand{\theequation}{S\arabic{equation}}
\renewcommand{\bibnumfmt}[1]{[S#1]}
\renewcommand{\citenumfont}[1]{S#1}
\addtolength{\textfloatsep}{5mm}

\section*{The variational parameters}
In Table \ref{tbl:parameters} we present the trial wave function parameters (Eq. (18)) 
optimized with VMC calculation for small $^4$He clusters $N\leq 50$.

\begin{table}[h]
\begin{center}
  \caption{
  The optimal parameters for the trial wave function (Eq. 18).
  \label{tbl:parameters}}
\vspace{0.3cm}
       {\renewcommand{\arraystretch}{1.25}%
\begin{tabular}
{c@{\hspace{4mm}} c@{\hspace{5mm}}  c@{\hspace{5mm}} c@{\hspace{5mm}} c}
\hline\hline 
   $N$ &       $p_0$     &     $p_1$        &        $p_2$    &    $p_5$ \\
\hline
2  &	1.	& 0.0309	& 0.3002	& 0.9022 \\
3  &	0.5	& 0.2136	& 0.1784	& 0.8884 \\
4  &	0.3333	& 0.2547	& 0.1774	& 0.8898 \\
5  &	0.25	& 0.2654	& 0.2975	& 0.877  \\
6  &	0.2	& 0.2279	& 0.1076	& 0.8825 \\
7  &	0.1667	& 0.2257	& 0.1166	& 0.8844 \\
8  &	0.1429	& 0.2063	& 0.1559	& 0.8822 \\
9  &	0.125	& 0.1972	& 0.1455	& 0.8829 \\
10 &	0.1111	& 0.1744	& 0.1202	& 0.8839 \\
15 &	0.0714	& 0.1379	& 0.1896	& 0.8823 \\
20 &	0.0714	& 0.1046	& 0.2401	& 0.881 \\
25 &	0.0417	& 0.093	        & 0.2526	& 0.8795 \\
30 &	0.0345	& 0.0796	& 0.2733	& 0.8789 \\
35 &	0.0294	& 0.07	        & 0.2727	& 0.8801 \\
40 &	0.1029	& 0.0405	& 0.3052	& 0.884 \\
45 &	0.0926	& 0.0364	& 0.3276	& 0.8808 \\
50 &	0.1068	& 0.0262	& 0.3326	& 0.8798 \\
\hline\hline
\end{tabular}}
\end{center}
\end{table}

\section*{The contacts}
In Table \ref{tbl:contacts} we present the contacts for small $^4$He clusters $N\leq 50$
calculated using a mixed VMC-DMC estimate.
The table includes contacts ${C}_n^{(N)}$ for $n=2-5$ 
coalescing particles.
As explained in the main body of this manuscript, the contacts were extracted from
the calculated $n$-particle densities $\rho_n^{(N)}(r)$. 

\begin{table}[h]
\begin{center}
  \caption{
  The numerical values of the $n$-body contacts in the $N$-body system ${C}_n^{(N)}$ 
  for $^4$He clusters in the range $N\in(2,50)$, calculated using the VMC-DMC mixed estimate.
  \label{tbl:contacts}}
\vspace{0.3cm}
       {\renewcommand{\arraystretch}{1.25}%
\begin{tabular}
{c@{\hspace{3mm}} c@{\hspace{4mm}}  c@{\hspace{4mm}} c@{\hspace{5mm}} c@{\hspace{5mm}} }
\hline\hline 
   $N$ &       $C_2^{(N)}$     &     $C_3^{(N)}$        &        $C_4^{(N)}$    &    $C_5^{(N)}$ \\
\hline
    2  & $  1.00e+00 \pm 0e+00$ & & &   \\ 
    3  & $  1.78e+01 \pm 3e-02$ & $  1.00e+00 \pm 0e+00$   & &  \\ 
    4  & $  4.88e+01 \pm 1e-01$ & $  6.55e+00 \pm 4e-03$   & $  1.00e+00 \pm 0e+00$ & \\ 
    5  & $  9.11e+01 \pm 3e-01$ & $  1.95e+01 \pm 4e-02$   & $  6.19e+00 \pm 6e-03$ & $  1.00e+00 \pm 0e+00$ \\ 
    6  & $  1.43e+02 \pm 6e-01$ & $  4.08e+01 \pm 9e-02$   & $  1.93e+01 \pm 8e-03$ & $  6.18e+00 \pm 1e-02$ \\ 
    7  & $  2.01e+02 \pm 8e-01$ & $  7.15e+01 \pm 2e-01$   & $  4.45e+01 \pm 1e-02$ & $  2.11e+01 \pm 4e-02$ \\ 
    8  & $  2.67e+02 \pm 1e+00$ & $  1.11e+02 \pm 3e-01$   & $  8.31e+01 \pm 8e-02$ & $  5.06e+01 \pm 2e-01$ \\ 
    9  & $  3.38e+02 \pm 2e+00$ & $  1.59e+02 \pm 6e-01$   & $  1.39e+02 \pm 9e-02$ & $  1.03e+02 \pm 5e-01$ \\ 
   10  & $  4.15e+02 \pm 2e+00$ & $  2.13e+02 \pm 7e-01$   & $  2.07e+02 \pm 5e-01$ & $  1.74e+02 \pm 1e+00$ \\ 
   15  & $  8.54e+02 \pm 6e+00$ & $  6.01e+02 \pm 3e+00$   & $  8.39e+02 \pm 2e+00$ & $  1.08e+03 \pm 1e+01$ \\ 
   20  & $  1.36e+03 \pm 1e+01$ & $  1.14e+03 \pm 8e+00$   & $  1.92e+03 \pm 3e+00$ & $  3.07e+03 \pm 3e+01$ \\ 
   25  & $  1.93e+03 \pm 2e+01$ & $  1.78e+03 \pm 1e+01$   & $  3.37e+03 \pm 6e+00$ & $  6.13e+03 \pm 6e+01$ \\ 
   30  & $  2.52e+03 \pm 2e+01$ & $  2.51e+03 \pm 2e+01$   & $  5.14e+03 \pm 9e+00$ & $  1.02e+04 \pm 1e+02$ \\ 
   35  & $  3.15e+03 \pm 3e+01$ & $  3.34e+03 \pm 3e+01$   & $  7.30e+03 \pm 9e+00$ & $  1.56e+04 \pm 2e+02$ \\ 
   40  & $  3.83e+03 \pm 4e+01$ & $  4.26e+03 \pm 4e+01$   & $  9.83e+03 \pm 6e+00$ & $  2.23e+04 \pm 2e+02$ \\ 
   45  & $  4.48e+03 \pm 5e+01$ & $  5.14e+03 \pm 5e+01$   & $  1.23e+04 \pm 1e+01$ & $  2.87e+04 \pm 3e+02$ \\ 
   50  & $  5.16e+03 \pm 6e+01$ & $  6.09e+03 \pm 6e+01$   & $  1.50e+04 \pm 1e+01$ & $  3.62e+04 \pm 4e+02$ \\ 
\hline\hline
\end{tabular}}
\end{center}
\end{table}

\begin{thebibliography}{99}
\bibitem{WerCas12a}
  Eqs. \eqref{momentum_distrib_1b}, \eqref{potential}, and \eqref{chi_r} 
  were derived by Werner and Castin in \cite{WerCas12ref}. 
  However, their definition of the 2-body contact
  $C=4\pi m/\hbar^2 dE/d(-1/a_s)$
  differs from ours. The two definitions coincide in the zero-range limit.
\bibitem{extract_contacts}
  To extract the contacts we have minimized the function 
  $I=\int_0^{r_{90}}|\rho_n^{(N)}-C_n^{(N)}\rho_n|^2dr$, choosing $r_{90}$
  such that 
  $\rho_n(r_{90})$ is $90\%$ of the $n$-body density peak $\rho_n(r_{peak})$, and 
  $r_{90}\leq r_{peak}$.
  We have found that this procedure is robust, as replacing $r_{90}$ by 
  $r_{85}$ or $r_{95}$ affects the contacts by less than $1\%$.
\end{thebibliography}
\end{document}